\documentstyle[epsfig]{elsart}
 
\newcommand{\La}{\Lambda}

\newcommand{\Si}{\Sigma}
\newcommand{\be}{\begin{eqnarray}}
\newcommand{\ee}{\end{eqnarray}}
\newcommand{\ri}{\rightarrow}

\begin{document}
\hspace{11cm}NT@UW-99-50 

\hspace{9cm}DOE/ER/40561-68-INT99
\begin{frontmatter}
 
\title{\bf The reactions $pp\rightarrow p\Lambda K^+$ and
$pp\rightarrow p\Sigma^0K^+$ near their thresholds }

\author{A. Gasparian$^{a,b}$, J. Haidenbauer$^a$, C. Hanhart$^c$,
L. Kondratyuk$^b$,}
\author{and J. Speth$^a$}

{\small $^a$Institut f\"{u}r Kernphysik, Forschungszentrum J\"{u}lich
GmbH,}\\ {\small D--52425 J\"{u}lich, Germany} \\
{\small $^b$Institute of Theoretical and Experimental Physics,} \\
{\small 117259, B.Cheremushkinskaya 25, Moscow, Russia}\\
{\small $^c$Nuclear Theory Group and INT, Dept. of Physics, 
University of Washington,}\\{\small  Seattle, 
WA 98195-1560, USA} \\

\begin{abstract}
The reactions $pp\rightarrow p\Lambda K^+$ and $pp\rightarrow p\Sigma^0K^+$ 
are studied near their thresholds. The strangeness production process
is described by the $\pi$- and $K$ exchange mechanisms. Effects from the
final state interaction in the hyperon-nucleon system are taken into
account rigorously. The $\Lambda$ production turns out to be 
dominated by $K$ exchange whereas $K$- as well as $\pi$ exchange 
play an important role for the $\Sigma^0$ case. It is shown that
the experimentally
observed strong suppression of $\Sigma^0$ production compared to 
$\Lambda$ production at the same excess energy can be explained 
by a destructive interference between $\pi$ and $K$ exchange
in the reaction $pp\rightarrow p\Sigma^0K^+$.
Implications of such an interference on the reaction 
$pp\rightarrow n\Sigma^+K^+$ are pointed out. 
\end{abstract}
\end{frontmatter}
 
Recently the total cross sections for the reactions $pp\rightarrow
p\Lambda K^+$ and $pp\rightarrow p\Sigma^0K^+$ were measured for the first 
time in the threshold region \cite{Bal,TOF,Sew}. 
Certainly the most interesting aspect of these new data is the observed 
strong suppression of the $\Sigma^0$ production in comparison to the
$\Lambda$ channel: 
at the same excess energy the cross section for the $\Sigma^0$
production is about a factor of 25 smaller than the one for
the $\Lambda$ production \cite{Sew}. 
This is indeed rather surprising, specifically because
data at higher energies \cite{Fla,Vig} indicate that the cross section for
$\Lambda$ production exceeds the one for $\Sigma^0$ production only
by a factor of around 2.5. 

\begin{figure}[h]
\begin{center}
\epsfig{file=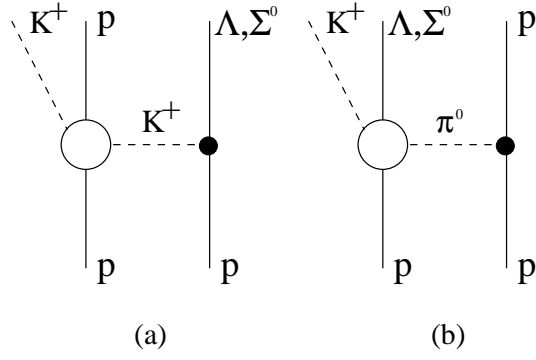, width=7.0cm}
\end{center}
\caption{Mechanisms for the reactions $pp\rightarrow p\Lambda K^+,
p\Sigma^0K^+$ considered in the present investigation: 
(a) kaon exchange; (b) pion exchange.}
\label{fig1}
\end{figure}

In principle, this strong suppression of the $\Sigma^0$ production 
compared to the $\Lambda$ case 
can be understood if one assumes that the hyperon production
is solely due to the $K$-exchange diagram, depicted in Fig.~\ref{fig1}(a).
In this case the same elementary ($K^+ p$) re-scattering amplitude 
enters in the two reactions. Therefore the ratio 
$\sigma_\Lambda/\sigma_{\Sigma^0} :=
\sigma_{pp\rightarrow p\Lambda K^+}/ \sigma_{pp\rightarrow p\Sigma^0K^+}$
will be given essentially by the ratio of the coupling constants at the 
vertices from which the $K$ meson emerges, i. e. by 
$g^2_{\Lambda NK}/g^2_{\Sigma NK} $. 
These coupling constants are not very well known experimentally.   
However, they can be inferred from SU(3) flavour symmetry - a symmetry which,  
so far, has been rather successfully employed in investigations
of reactions involving hyperons. Specifically according to SU(6) 
this ratio is 27 \cite{SU6}, a value which coincides almost 
exactly with the experimental cross section ratio. 

None-the-less, already a simple estimation 
of the elementary scattering processes ($K^+p \rightarrow 
K^+p$ and $\pi^0 p \rightarrow K^+\Lambda, K^+\Sigma^0$, respectively)  
based on experimental amplitudes reveals that the contribution of 
the $\pi$-exchange 
diagram (Fig.~\ref{fig1}(b)) to hyperon production should be not negligible, 
cf. Ref.~\cite{Sew}. Strong evidence for the relevance of $\pi$--exchange 
comes also from more detailed model 
calculations \cite{Lag,Li,Tsu,Sib,Kai,Shyam}. 
(It should be said, however, that most of these studies focus on data 
at rather large excess energies or look at the $\Lambda$ channel
only.) Indeed the $\Lambda/\Sigma^0$ production ratio 
estimated in Ref.~\cite{Sew}, considering $K$- as well as $\pi$
exchange, is roughly 3.6, i.e. about a factor of 8 below the measurement. 

Because of this situation another explanation for the observed 
suppression of the $\Sigma^0$ production was suggested in Ref.~\cite{Sew},
namely effects from the strong $\Sigma N$ final state interaction (FSI) 
leading to a $\Sigma N \rightarrow \Lambda N$ conversion. (Note that
such FSI effects have been ignored in the above discussion altogether!) 
Evidence suggesting this conversion hypothesis can be readily found in the
literature. E.g., the predictions of modern meson-theoretical models
of the hyperon-nucleon $YN$ interaction for the $\Lambda N$ cross 
section show a large cusp structure at the $\Sigma N$ threshold, which
arises from the strong coupling between the $\Lambda N$
and $\Sigma N$ channels in those models \cite{Hol,Nij}. 
Inclusive measurements of $K^+$ production in the reaction 
$pp \rightarrow K^+ X$ at 2.3 GeV show a significant enhancement
near the $\Sigma N$ threshold
\cite{Sie}. Finally, data on the reaction $K^- d \rightarrow \pi^- 
\Lambda p$ show a sharp peak at an effective mass of $m_{\Lambda p}  
\approx$ 2130 $MeV/c^2$, i.e. at the $\Sigma N$ threshold \cite{Tan}. 
(Cf. also corresponding theoretical investigations in Refs.
\cite{Kud,Del}.) Thus, it is obvious
that there is a strong enhancement of the $\Lambda$ counting rate in
those reactions. However, it is much less clear whether this 
enhancement is indeed due to produced ``real'' $\Sigma$'s being converted into 
$\Lambda$'s in the FSI so that the number of experimentally observed 
$\Sigma$'s in any exclusive measurement will be greatly reduced. 

In the present paper we want to study the $\Lambda$ and $\Sigma^0$
production cross section in $pp$ collisions near threshold. Thereby
our emphasis will be on a careful treatment of the FSI. 
Specifically we want to investigate in detail possible effects from
the $\Sigma \leftrightarrow \Lambda$ conversion. In order to have
a solid basis for our study we employ microscopic $YN$ interaction models
from the J\"ulich \cite{Hol} and Nijmegen \cite{Nij} groups. 
The J\"ulich $YN$ model is derived in the meson-exchange picture 
and has been constructed according to the same guidelines as those used 
in the Bonn $NN$ potential \cite{Bonn}.
The model is given in momentum space and contains the full nonlocal 
structure resulting from the relativistic meson exchange framework.
The parameters at the $NN$ vertices (coupling constants and cutoff 
masses in the form factors) are taken from the Bonn potential.
The coupling constants at the strange vertices are determined from 
$SU(6)$ symmetry relations. The only free parameters in this model are
the cut-off masses of the form factors at the strange vertices - which
are determined by a fit to the empirical hyperon-nucleon data.
The model describes existing $\Lambda N$ and $\Sigma N$ 
observables reasonably well as can be seen in Ref.~\cite{Hol}. 

Besides the J\"ulich model we will also employ potential models 
provided by the Nijmegen group \cite{Nij}. This will allow us
to investigate the model dependence of our results. The Nijmegen
NSC97 potentials provide a comparably good description of the available
$YN$ data. However, there are quite significant differences in the
dynamical input of the Nijmegen models and that of the J\"ulich 
group as discussed, e.g., in Ref.~\cite{Reu}.
Thus, it is possible that these differences play a role in the
present analysis, specifically because in the associated strangeness
production it will be mostly the off-shell properties of these
$YN$ models that enter into the calculation. 

\begin{figure}[h]
\begin{center}
\epsfig{file=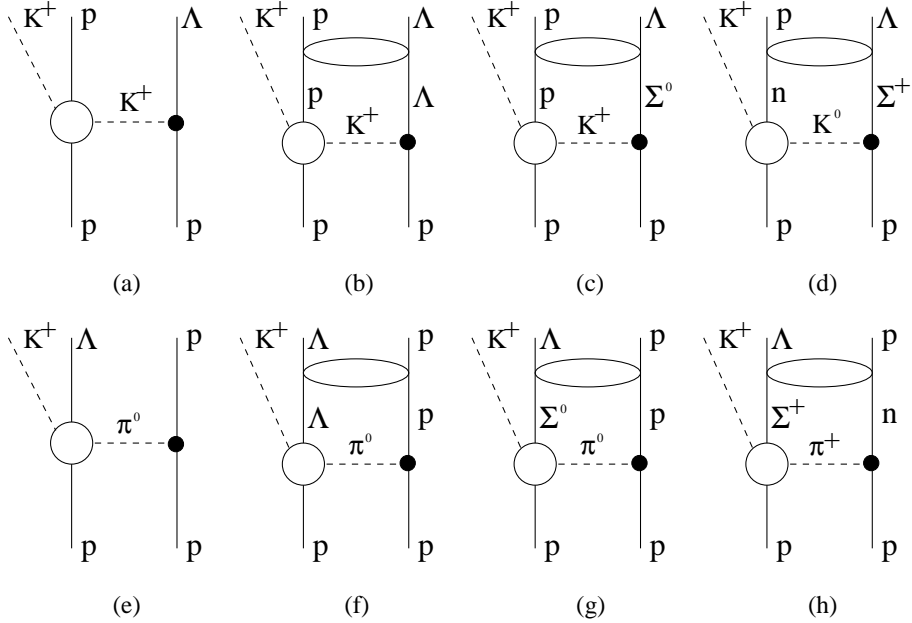, width=12.0cm}
\end{center}
\caption{Contributions to the production amplitude for 
$pp\to pK^+\Lambda$ when the final-state interaction is
included, cf. Eq.~(\ref{DWBA}). The open circles and ellipses
stand for T-matrices. }
\label{fig2}
\end{figure}

We treat the associated strangeness production in the standard
distorted wave Born approximation. Thus, the production amplitude
$M$ is obtained from the formal equation
\begin{equation}
 M \ = \ A \ + \ A G_0 T_{YN} \ , 
\label{DWBA} 
\end{equation}
where $A$ is the elementary production process ($\pi$- and/or $K$ 
exchange, Fig.~\ref{fig1}), $T_{YN}$ the interaction in the 
final state, and $G_0$
the free $YN$ propagator. Note that the second term on the right-hand
side involves, in fact, a sum over the (coupled) $\Lambda N$ and 
$\Sigma $N states. This can be seen from the graphic representation
of Eq.~\ref{DWBA} in Fig.~\ref{fig2} (for the case of $\Lambda$
production). 

Following standard rules \cite{bjo} one gets the following 
expression for the antisymmetrized Born amplitudes,
\be
\nonumber
A^{\mu_1\mu_2\mu_3\mu_4}_K(\vec{p}_1,\vec{p}_2 ;\vec{p}_3,\vec{p}_4)
&=&F_{K}(\vec{p}_3-\vec{p}_1)\Gamma_{NYK}^{\mu_1\mu_3}\frac{2\omega_K}
{k_K^2-m_K^2}T_{KN}^{\mu_2\mu_4} \\
&-&(\vec{p}_1\leftrightarrow\vec{p}_2,\mu_1\leftrightarrow \mu_2) \\
\nonumber
A^{\mu_1\mu_2\mu_3\mu_4}_\pi(\vec{p}_1,\vec{p}_2 ;\vec{p}_3,\vec{p}_4)
&=&F_{\pi}(\vec{p}_4-\vec{p}_1)\Gamma_{NN\pi}^{\mu_1\mu_4}
\frac{2\omega_{\pi}}{k_{\pi}^2-m_{\pi}^2}T_{\pi N\ri KY}^{\mu_2\mu_3} \\
&-&(\vec{p}_1\leftrightarrow\vec{p}_2,\mu_1\leftrightarrow \mu_2)
\ee
where the $\mu_i$ ($\vec p_i$) are the spin projections (momenta)
of the baryons, 
$k_K$ and $k_{\pi}$ are the four momenta of the exchanged kaon and pion,
and $\omega_K=\sqrt{(\vec{p}_3-\vec{p}_1)^2+m_K^2}$ and 
$\omega_{\pi}=\sqrt{(\vec{p}_4-\vec{p}_1)^2+m_{\pi}^2}$ are their energies.
$T_{KN}$ and $T_{\pi N\ri KY}$ are the amplitudes of the corresponding 
elementary reactions. $\Gamma_{NYK}$ and $\Gamma_{NN\pi}$ are 
vertex functions for the corresponding baryon-baryon-meson 
vertices, 
\begin{eqnarray}
\Gamma^{\mu_i\mu_j}=&&i\frac{f_{B_iB_jM}}{m_M}\frac{1}{(2\pi)^{3/2}}\frac{1}
{\sqrt{2\omega_M}}N_iN_j \nonumber \\ 
&& \ \ \times \ \chi^{\dagger}_{\mu_j} \vec{\sigma}
\cdot
\left[ \vec k_M - 
\omega_M  \left(\frac{\vec p_i}
{\epsilon_{p_i}}+\frac{\vec
p_j}{\epsilon_{p_j}}\right) +
\left(\frac{{\vec{p}_i}{}^2\vec{p}_j-{\vec{p}_j}{}^2\vec{p}_i}
{\epsilon_{p_i}\epsilon_{p_j}}\right) \right] \chi_{\mu_i}
\end{eqnarray}
where $\epsilon_{p_i}=E_{p_i}+m_i$ and 
$N_i=\sqrt{\frac{\epsilon_{p_i}}{2E_{p_i} }}$,
and $F_K$ and $F_{\pi}$ are form factors which are assumed to
be of monopole type, i.e. 
$F_M(\vec{k}) = (\Lambda^2_M - m^2_M)/(\Lambda^2_M+\vec{k}^2)$. 

The total cross section is obtained from 
\begin{equation}
\sigma = \frac{1}{4v}\sum_{\mu_i,\mu_f} \int
d^3p_3d^3p_4d^3p_{K^+}(2\pi)^
4\delta^{(4)}(P_f-P_i)|M_{i\rightarrow f}|^2
\label{cros}
\end{equation}
where $p_K^+$ is the momentum of the produced kaon and
$v$ is the relative velocity of the two initial protons. 

The vertex parameters employed in the present study are compiled
in Table~\ref{tab:par}. For the calculation with the J\"ulich $YN$
interaction we take the same values that were used in this model.
In case of the Nijmegen model we take over only their coupling
constants but not their vertex form factors. For simplicity
reasons we use here also a monopole form - but with a uniform 
cutoff parameter of $\Lambda$ = 1.3 GeV. 

The elementary amplitudes $T_{KN}$ and $T_{\pi N \ri KY}$
can be taken from microscopic models of $KN$ scattering \cite{Mon1} 
and of the reaction $\pi N \rightarrow K\Lambda, K\Sigma$ \cite{Mon2} 
that were developed by our group. 
However, since in the present more exploratory study we would like to
focus mainly on the FSI effects we will restrict ourselves
to a simplified treatment of the production amplitude. Thus,
instead of the full (off-shell) $KN$ and $\pi N \rightarrow KY$
transition amplitudes we use the scattering length and on-shell
threshold amplitudes of those reactions. The off-shell extrapolation
of the amplitudes is done by multiplying those quantities with
the same form factor that is used at the vertex where the 
exchanged meson is emitted. Furthermore we take into account only
s waves. 
For the $KN$ s-wave scattering lengths 
we employ the values $a^0 = -0.038 \ fm$ and $a^1 = -0.304 \ fm$
(for the isospin 0 and 1 states) resulting from our $KN$ model \cite{Mon1}, 
which are in good agreement with experimental information \cite{akn}. 
For the $\pi N \rightarrow K Y$ amplitudes we use 
$f_{\pi^0 p\ri K^+\La}=(-0.06+{\it i}0.48) \times 10^{-1} \ fm$,
$f_{\pi^0 p\ri K^+\Si^0}=(0.45-{\it i}0.35) \times 10^{-1} \ fm$, and
$f_{\pi^+ p\ri K^+\Si^+}=(-0.11+{\it i}0.21) \times 10^{-1} \ fm$.
The value for the first reaction yields 
$|f_{\pi^0 p\ri K^+\La}|^2=29\ \mu b/sr$ which is comparable to
the number deduced by F\"aldt and Wilkin \cite{FW} from $\pi^- p$ 
data \cite{pim}. For the other reactions we get 
$|f_{\pi^0 p\ri K^+\Si^0}|^2=33 \ \mu b/sr$ and
$|f_{\pi^+ p\ri K^+\Si^+}|^2=5.7 \ \mu b/sr$.
>From those scattering amplitudes the T-matrices are obtained
via the relation 
\be
f_{MB\ri M'B'} = -4\pi^2 \frac{\sqrt{E_B\omega_ME_{B'}\omega_{M'}}}
{E_B+\omega_M} T_{MB\ri M'B'} 
\ee
using the threshold kinematics. 

Since we will concentrate on energies very close to the thresholds 
we consider only the lowest partial waves in the outgoing channels;
the $\Lambda p$ and $\Sigma^0 p$ system can be in an $^1S_0$ or in a
$^3S_1$ state and the $K^+$ is assumed to be in an s-wave relative
to the $YN$ state. Angular momentum and parity conservation then
tells us that the initial $pp$ system has to be in the $^3P_0$ or in 
the $^3P_1$ state. Note that the $^3S_1$ partial wave couples to the
$^3D_1$ and this coupling is taken into account in our calculations.

We do not take into account the initial state interaction (ISI)
between the protons. 
Based on a recent examination of the influence of the ISI for the 
reaction $pp \rightarrow pp\eta$ by Batini\'c et al. \cite{Bat} 
we expect that the neglection of the ISI should result in an 
overestimation of the cross sections by a factor of around 3 in our
calculation (cf. also Ref.~\cite{HaKa}). But since the thresholds for 
the $\Lambda$ and $\Sigma^0$ production are relatively close together 
(at $T_{lab}$ = 1582 MeV and at $T_{lab}$ = 1796 MeV) and, 
moreover, the energy dependence of the $NN$ interaction is relatively
weak in this energy region we expect that the ISI effects are very
similar for the two strangeness production channels and therefore
should roughly drop out when ratios of the cross sections are taken. 
Thus, we believe that our model calculation allows a quite reliable
estimation of $\sigma_\Lambda/\sigma_{\Sigma^0}$. The same is also 
true for a comparison of the relative magnitude of the pion- and 
kaon-exchange contributions. 

The FSI increases considerably the number of contributing amplitudes
as can be seen from the graphs shown in Fig.~\ref{fig2}.
Besides the Born terms (Fig.~\ref{fig2}a,e)
there are contributions from the "diagonal" FSI (Fig.~\ref{fig2}b,f)
and from the transitions with $p \Sigma^0$ and $n \Sigma^+$ 
intermediate states (Fig.~\ref{fig2}c,g and d,h, respectively). 
All these contributions have to be added coherently for the
evaluation of the production cross section. 
In order to investigate the influence of the FSI in detail and 
to clarify the roles played by the $\pi$- and $K$-exchange we
have also evaluated the contributions of the individual diagrams 
and compiled them in Tables \ref{tab:J} (for the J\"ulich model A) 
and \ref{tab:N} (for the Nijmegen model NSC97f). The analysis is done
for the data at the highest available excess energy, i.e. 
13.2 MeV for the $\Lambda$ production and 13.0 MeV for the
$\Sigma^0$ production.  

Let us first discuss the $K$ exchange. The cross section ratio
resulting from the Born diagram alone is around 16 for the J\"ulich model,
cf. Table~\ref{tab:J}. Based on the ratio of the coupling constants 
(Table~\ref{tab:par}) one would have expected a value close to 27.
However, one has to keep in mind that a much harder form factor
is employed at the $\Sigma N K$ vertex than at the $\Lambda N K$
vertex (cf. Ref.~\cite{Hol}) which obviously has a strong impact
on the actual results. Now we add the
amplitude of the "diagonal" FSI which means diagram (b) of 
Fig.~\ref{fig2} for $\Lambda$ production and diagrams with
$\Sigma^+ n$ and $\Sigma^0 p$ intermediate states for $\Sigma^0$
production. This is the step where a possible conversion 
effect $\Sigma N \rightarrow \Lambda N$ should become visible.
The employed $\Sigma N$ t-matrix is the solution of the
coupled-channel scattering equation and therefore includes
the flux going from the $\Sigma N$ to the $\Lambda N$ system. 
Indeed the consideration of the "diagonal" 
FSI enhances the cross section of the $\Lambda$ 
channel and reduces the one of the $\Sigma^0$ channel. 
As a consequence, the resulting cross section ratio 
becomes significantly larger than the value obtained from the Born
term and is even close to the experimental value, cf. Table~\ref{tab:J}.
In the final step we add the diagram where a transition 
$\Sigma N \leftrightarrow \Lambda N$ occurs in the FSI (cf, 
Fig.~\ref{fig2}c,d). This leads to further modifications in the cross 
sections and to a further increase in cross section ratio.

In case of pion exchange the Born diagrams yield a cross
section ratio of 0.9. This value is somewhat
larger than the estimate presented in Ref.~\cite{Sew} 
because now the isospin $3 \over 2$ component
of the $\pi N \rightarrow K \Sigma$ amplitude was taken
into account as well. Adding the FSI step by step 
increases the cross section ratio somewhat, but it remains
far below the experiment. 

Thus, it's clear that, in principle, $K$ exchange alone
can explain the cross section ratio - especially after 
inclusion of FSI effects. 
However, we also see from Table~\ref{tab:J} that $\pi$
exchange definitely yields a significant contribution 
to the $\Sigma^0$ channel and therefore it cannot be
ignored. Indeed, the two production mechanisms
play quite different roles in the two reactions under
consideration, cf. Table~\ref{tab:J}. 
$K$ exchange yields by far the dominant contribution for
$pp\rightarrow p\Lambda K^+$. Here the cross section obtained
from $\pi$ exchange is about an order of magnitude smaller.
In case of the reaction $pp\rightarrow p\Sigma^0K^+$,
however, $\pi$- and $K$ exchange give rise to contributions
of comparable magnitude. This feature becomes very important when
we now add the two contributions coherently and consider different
choices for the relative sign between the $\pi$ and $K$ exchange
amplitudes. In one case (indicated by ``$K+\pi$'' in Table~\ref{tab:J})
the $\pi$ and $K$ exchange contributions
add up constructively for $pp\rightarrow p\Sigma^0K^+$ and
the resulting total cross section is significantly larger than
the individual results. For the other choice (indicated by
``$K-\pi$'') we get a destructive interference between the amplitudes
yielding a total cross section that is much smaller.
Consequently, in the latter case the cross section ratio is
much larger and, as a matter of facts, in rough agreement with the
experiment (cf. Table~\ref{tab:J}). 
 
In this context it is interesting to look at corresponding results for 
the reaction $pp\rightarrow n\Sigma^+K^+$. At the excess energy of 
13 MeV the predicted cross sections are 86 (``$K+\pi$'') and
229 $nb$ (``$K-\pi$''), respectively. Thus, the interference 
pattern is just the opposite as for $pp\rightarrow p\Sigma^0K^+$,
cf. Table~\ref{tab:J}. 
For the ``$K-\pi$'' case favoured by the experimental 
$\sigma_\Lambda/\sigma_{\Sigma^0}$ ratio our calculation yields a cross 
section for $pp\rightarrow n\Sigma^+K^+$ that is about 3 times 
larger then the one for $pp\rightarrow p\Sigma^0K^+$. Such a ratio
is in fair agreement with data and model calculations at higher
energies, see, e.g. Ref.~\cite{Lag}. The other choice, ``$K+\pi$'', 
leads to a $\sigma_{\Sigma^+}$ that is a factor of about 3 smaller than
$\sigma_{\Sigma^0}$ - a result which is rather difficult to reconcile
with the present knowledge about these reactions at higher energies. 
Obviously it would be very interesting to determine also
the ratio $\sigma_{\Sigma^+}/\sigma_{\Sigma^0}$ close to threshold.
It could be measured at, e.g., the COSY facility in J\"ulich \cite{COSY}. 

The results based on the Nijmegen model NSC97f \cite{Nij}
are compiled in Table~\ref{tab:N}. It is evident that the actual values 
of the cross sections as well as for the ratios are 
rather different from those obtained with the J\"ulich $YN$ interaction. 
Most strikingly the FSI (i.e. conversion effects) no longer leads to an
enhancement of the cross section ratio in case of $K$ exchange but 
to a reduction. Consequently, neither $K$- nor $\pi$ exchange lead to 
a ratio anywhere near to the experimental value. 
On a qualitative level, however, the results are still very similar. 
Again $K$ exchange is the dominant production mechanism for the
reaction $pp\rightarrow p\Lambda K^+$, whereas for 
$pp\rightarrow p\Sigma^0 K^+$ $K$- as well as $\pi$ exchange yield
contributions of comparable magnitude. Thus, like for the J\"ulich
model, a large cross section ratio $\sigma_\Lambda/\sigma_{\Sigma^0}$ 
can only be achieved if there is a destructive interference between
the $K$- and $\pi$ exchange in the reaction 
$pp\rightarrow p\Sigma^0 K^+$ (though such an interference does
not occur anymore for the same specific phase between the $K$- and $\pi$ 
exchange as chosen for the J\"ulich model, cf. Table~\ref{tab:N}).

We have also carried out calculations utilizing the other $YN$ models 
presented in Ref.~\cite{Nij} (NSC97a-e). The pertinent results 
exhibit again strong variations in the details. But qualitatively they 
are similar to the ones discussed above and therefore confirm our
findings. Thus we refrain from showing them here explicitly. 

Evidently, the results for the cross sections as well as 
of the ratios depend significantly on the employed FSI. Accordingly, 
the good agreement of the predictions based on the J\"ulich $YN$ model 
A with the experiment is certainly accidental.  
But the important message following from our investigation is
that only a destructive interference between $\pi$ and $K$ exchange 
can yield fairly large cross section ratios and therefore does offer 
a possible explanation for the experiment. If the two production
mechanisms add up constructively there is little chance of ever 
coming anywhere close to the experimental value of the ratio, 
as can be seen from the Tables. We should emphasize, however, that
our findings are based on the assumption that 
$\pi$ and $K$ exchange are the dominant mechanisms for associated
strangeness production. This postulate is certainly not 
unreasonable as demonstrated by the success of earlier investigations 
on the reaction $pp\rightarrow p\Lambda K^+$ \cite{Lag,Li,Sib}.
But it is conceivable
that other production mechanisms like the direct production or
the exchange of heavier mesons, specifically of the vector
mesons $\rho$ and $K^*$, play a role as well \cite{Tsu,Kai}. 
This, of course,
leads to quite a different scenario and it is certainly
desirable to carry out investigations in this direction in the future.

\begin{figure}[h]
\begin{center}
\epsfig{file=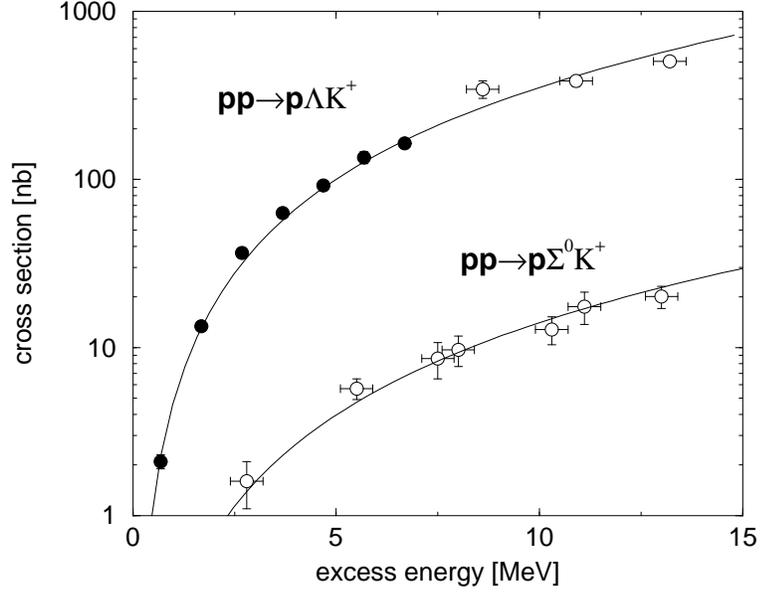, width=10cm}
\end{center}
\caption{Total cross sections for the reactions 
$pp\rightarrow p\Lambda K^+$ and $pp\rightarrow p\Sigma^0K^+$ 
employing the J\"ulich $YN$ model for the FSI. 
The shown results correspond to the choice ``$K-\pi$'' for the
relative sign, cf. text. The curves for 
$pp\rightarrow p\Lambda K^+$ and $pp\rightarrow p\Sigma^0K^+$ 
are normalized by a factor of 0.25 and 0.30, respectively, 
in order to account for effects of the inital state interaction - 
as described in the text.
The experimental data are from Refs.~\protect\cite{Bal}
(filled circles) and \protect\cite{Sew} (open circles).}
\label{fig3}
\end{figure}

In Fig.~\ref{fig3} we show the total cross sections as a function of 
the excess energy for the ``$K-\pi$'' case and with the J\"ulich $YN$ 
model as the FSI. It is clear from Table~\ref{tab:J} that our
model calculations overestimates the absolute values of the
empirical cross sections by roughly a factor 4. This is not too
surprising because, as already mentioned earlier, effects from
the ISI are not taken into account. Therefore, we normalized 
the results by the factors 0.25 ($pp\rightarrow p\Lambda K^+$) and 0.30
($pp\rightarrow p\Sigma^0K^+$), respectively. Then it can be easily
seen that the model calculations yield an energy dependence that
is in rather nice agreement with the experiment. This suggests that,
like in the case of pion production \cite{Mil,HoM}, the energy dependence
of the cross section is primarily influenced by the FSI between
the baryons and that effects from other possible FSI's (in the 
$KN$ and/or $KY$ systems) play a minor role. 

In summary, we have studied the reactions $pp\rightarrow p\Lambda K^+$ 
and $pp\rightarrow p\Sigma^0K^+$ near their thresholds. 
The strangeness production process is described by the $\pi$- and 
$K$ exchange mechanisms. Effects from the
final state interaction in the hyperon-nucleon system are taken into
account rigorously. 
Our study suggests that the $\Lambda$ production is dominated by 
$K$ exchange whereas $K$- as well as $\pi$ exchange 
play an important role for the $\Sigma^0$ case. 
Furthermore we found that the experimentally
observed large cross section ratio $\sigma_{pp\rightarrow p\Lambda K^+}
/\sigma_{pp\rightarrow p\Sigma^0K^+}$ 
(at the same excess energy) of around 25 cannot be 
explained by FSI effects. Rather we conclude from our
investigation that a destructive interference between $\pi$ and 
$K$ exchange in the reaction $pp\rightarrow p\Sigma^0K^+$ could be
the origin of the strong suppression of $\Sigma^0$ production.

\vskip 1cm 

{\bf Acknowledgments}

We acknowledge communications with A. Deloff and J.M. Laget concerning
their model calculations of the associated strangeness production. 
C.H. is grateful for the financial support through a Feodor-Lynen
Fellowship of the Alexander-von-Humboldt Foundation 
(DE-FG03-97ER41014).
This work was also supported by the INTAS grant no. 96-0597.

\vfill \eject 

\begin{table}[h]
\caption{Vertex parameters used in the present calculation.
In case of the J\"ulich $YN$ interaction the same values are used 
as in the model \cite{Hol}. For the Nijmegen potential only the 
coupling constants are taken over from Ref.~\cite{Nij}.}
\vskip 0.5cm
\begin{center}
\begin{tabular}{|c|c|c|c|c|c|c|}
\cline{2-7}
\multicolumn{1}{c|}{ }& \multicolumn{3}{c}{J\"ulich model A}& 
\multicolumn{3}{|c|}{Nijmegen model NSC97f}\\
\hline
$Vertex$ & $g/\sqrt{4\pi}$ & $f/\sqrt{4\pi}$ & $\La\ [GeV]$ & $g/\sqrt{4\pi}$ &
 $f/\sqrt{4\pi}$ & $\La\ [GeV]$ \\
\hline
$NN\pi$ &  3.795&0.282& 1.3&3.671 &0.273 &1.3\\
\hline
$N\La K$ & -3.944 &-0.268& 1.2&-4.925 &-0.335 &1.3\\
\hline
$N\Si K$ & 0.759 &0.0497& 2.0& 1.501& 0.0983 &1.3\\
\hline
\end{tabular}
\end{center}
\label{tab:par}
\end{table}

\vskip 1cm 

\begin{table}[h!!!]
\caption{Contributions of different diagrams to the total cross
section of the reactions $pp\rightarrow p\Lambda K^+,p\Sigma^0K^+$.
The results for $\La$ ($\Si^0$) production are for the excess energy of
13.2 MeV (13.0 MeV). The J\"ulich $YN$ model A \cite{Hol}
is employed for the final-state interaction.
The indication for the diagrams refers to Fig.~\ref{fig2}.
In case of ``diagonal'' only the diagonal channel is included in 
the intermediate state, i.e. only $p\La$ for $\La$ 
production and $p\Si^0$ and $n\Si^+$ for $\Si^0$ production.}
\vskip 0.5cm
\begin{center}
\begin{tabular}{|c|c|c|c|c|}
\cline{2-5}
\multicolumn{1}{c|}{ }&diagrams &   $\sigma_{pp \rightarrow p \Lambda K^+}\  
[nb]$ &  $\sigma_{pp \rightarrow p \Sigma^0 K^+}\  [nb]$ & 
$\frac{\sigma_{pp \rightarrow p \Lambda K^+}}
{\sigma_{pp \rightarrow p \Sigma^0 K^+}}$ \\
\cline{1-5}
&(a)  & 739 & 46 & 16 \\
\cline{2-5}
$K$&``diagonal''   & 1113 & 34 & 33 \\
\cline{2-5}
&(a)-(d)   & 2426 &  57 & 43 \\
\cline{1-5}
\noalign{\vskip 0.02cm}
\hline
&(e)   & 71 & 77 & 0.9 \\
\cline{2-5}
$\pi$&``diagonal''  & 113 & 104 & 1.1 \\
\cline{2-5}
&(e)-(h)  & 113 & 105 & 1.1 \\
\hline
\noalign{\vskip 0.02cm}
\hline
$"K+\pi"$& all  & 2471 & 251 & 9.9 \\
\cline{1-5}
$"K-\pi"$& all  & 2607 & 73 & 36 \\
\cline{1-5}
\multicolumn{2}{|c|}{ experiment} & 505 $\pm$ 33 & 20.1 $\pm$ 3.0 & 25 $\pm$
6 \\
\hline
\end{tabular}
\label{tab:J}
\end{center}
\end{table}

\vfill \eject

\begin{table} 
\caption{Same as in Table~\ref{tab:J} employing the Nijmegen $YN$ model 
NSC97f \cite{Nij} for the final-state interaction.}
\vskip 0.5cm
\begin{center}
\begin{tabular}{|c|c|c|c|c|}
\cline{2-5}
\multicolumn{1}{c|}{ }&diagrams &   $\sigma_{pp \rightarrow p \Lambda K^+}\  
[nb]$ &  $\sigma_{pp \rightarrow p \Sigma^0 K^+}\  [nb]$ & 
$\frac{\sigma_{pp \rightarrow p \Lambda K^+}}
{\sigma_{pp \rightarrow p \Sigma^0 K^+}}$ \\
%\cline{1-5}
\hline
&(a)  & 1598 & 116 & 14 \\
\cline{2-5}
$K$&``diagonal'' & 1207 & 109 & 11 \\
\cline{2-5}
&(a)-(d) & 1367 & 209 & 6.5 \\
\hline
\noalign{\vskip 0.02cm}
\hline
&(e)   & 67 & 73 & 0.91 \\
\cline{2-5}
$\pi$&``diagonal'' & 63 & 190 & 0.33 \\
\cline{2-5}
&(e)-(h) & 63 & 219 & 0.29 \\
\hline
\noalign{\vskip 0.02cm}
\cline{1-5}
$"K+\pi"$& all  & 1460 & 432 & 3.4 \\
\cline{1-5}
$"K-\pi"$& all  & 1400 & 424 & 3.3 \\
\cline{1-5}
\multicolumn{2}{|c|}{ experiment} & 505 $\pm$ 33 & 20.1 $\pm$ 3.0 & 25 $\pm$
6 \\
\hline
\end{tabular}
\end{center}
\label{tab:N}
\end{table}

\phantom{x}

\vfill \eject 

\vskip 3cm 

\vfill \eject

\end{document}